\documentclass{article}
\usepackage{ijcai22}

\usepackage{times}
\usepackage{soul}
\usepackage{dpr}
\usepackage{fec}
\usepackage{url}
\usepackage{hyperref}
\usepackage[utf8]{inputenc}
\usepackage[small]{caption}
\usepackage{graphicx}
\usepackage{amsmath}
\usepackage{amsthm}
\usepackage{booktabs}
\usepackage{algorithm}
\usepackage[noend]{algorithmic}
\usepackage{subfigure}
\usepackage{multirow}
\usepackage{amsmath,mathabx}
\usepackage{pifont}
\usepackage{amssymb}
\usepackage{verbatim}
\usepackage{color}

\newcommand\sbullet[1][.5]{\mathbin{\vcenter{\hbox{\scalebox{#1}{$\bullet$}}}}}
\newcommand{\alagcro}{BufferSearch}
\newcommand{\ie}{\textit{i.e.}}
\newcommand{\eg}{\textit{e.g.}}
\newcommand{\imdb }{\text{IMDB}}
\newcommand{\mr}{\text{MR}}
\newcommand{\yelp}{\text{Yelp}}
\newcommand{\lstm}{\text{WordLSTM}}
\newcommand{\cnn}{\text{WordCNN}}
\newcommand{\bert}{\text{BERT}}
\newcommand{\xmark}{\ding{55}}%

\title{BufferSearch: Generating Black-Box Adversarial Texts With Lower Queries}

\author{
Wenjie Lv$^1$
\and
Zhen Wang$^1$\and
Yitao Zheng$^1$\and
Zhehua Zhong$^1$\and
Qi Xuan$^2$\And
Tianyi Chen$^3$
\affiliations
$^1$Hangzhou Dianzi University\\
$^2$Zhejiang University of Technology\\
$^3$Microsoft\\
\emails
\{211270023,wangzhen\}@hdu.edu.cn,
xuanqi@zjut.edu.cn,
tiachen@microsoft.com
}

\begin{document}

\maketitle
\begin{abstract}
Machine learning security has recently become a prominent topic in the natural language processing (NLP) area. The existing black-box adversarial attack suffers prohibitively from the high model querying complexity, resulting in easily being captured by anti-attack monitors. Meanwhile, how to eliminate redundant model queries is rarely explored. In this paper, we propose a query-efficient approach~\alagcro{}
to effectively attack general intelligent NLP systems with the minimal number of querying requests. In general, \alagcro{} makes use of historical information and conducts statistical test to avoid incurring model queries frequently. Numerically, we demonstrate the effectiveness of \alagcro{} on various benchmark text-classification experiments by achieving the competitive attacking performance but with a significant reduction of query quantity. Furthermore,~\alagcro{} performs multiple times better than competitors within restricted query budget. Our work establishes a strong benchmark for the future study of query-efficiency in NLP  adversarial attacks. 
\end{abstract}

\section{Introduction}
In the recent years, Deep Neural Networks (DNNs) have been shown to be effective in a variety of  natural language processing  applications, such as machine translation, fake news detection, question answering, and sentiment analysis~\cite{machine_translation_0,minaee2021deep,zhong2023mat}. Meanwhile, the security of DNNs is receiving increasing attentions, since they have been proved to be easily confused by adversarial samples by slightly perturbing the original input~\cite{759851e20d2e47aaad2a560211f6a126}. 

In general, there are two ways to fool NLP models via adversarial attack ,~\ie, white-box and black-box attack~\cite{zhang2021argot}. The white-box attack can access almost every aspects of the target model including the training dataset, architecture and parameters, which are not transparent to the black-box settings instead~\cite{2021aaaiFast}. In practice, black-box attack is more applicable since 
such inherent underlying information may be not accessible. However, the success of the existing state-of-the-art black-box attackers highly relies on querying the victim model sufficient many times~\cite{yoo-etal-2020-searching,Explain2Attack}. Such massive model queries might be easily detected by the anti-attack security system thereby still not convenient in reality. 

How to achieve efficient black-box NLP attack with the minimal number of model queries now becomes an open and hot problem. In fact, the heavy model queries derive from the two main stages of NLP black-box attacks,~\ie, one stage to calculate and rank word salience followed by another stage to employ prescribed transformation onto the important words. Both stages contribute a fair amount of model queries. To achieve high query efficiency, there are two recent works concentrating on the first stage to optimize the important word selection. In particular, E2A~\cite{Explain2Attack} trains an interpretable agent model to learn word salience scores by using a substitute training corpora within the similar fields of the target model. \cite{hossam2021improved} further improves the training process
in E2A~\cite{Explain2Attack}.  However, to the best of our knowledge, there is no work to eliminate redundant model queries on the second stage~\ie, perform word transformation on the important tokens.

On the other hand, in the related computer vision (CV) field, there exist a few works that establish Bayesian attackers to reduce query complexity by leveraging the historical information ~\cite{zhao2019design,ru2019bayesopt}, wherein they assume a Gaussian distribution to the perturbation variable and determine its hype-parameters via Bayesian optimization. However, these approaches cannot be easily extended to NLP tasks because of the discreteness of the perturbation space.  

In this paper, we study how to optimize the model query over the second stage (word transformation stage) and propose a~\alagcro{} to effectively attack NLP applications with significant amount of model query reduction. Our main contributions are summarized as follows. 

\noindent\textbf{Algorithmic Design.} We are the first work to optimize model queries of the word transformation stage in the general black-box NLP adversarial attacks.  The proposed~\alagcro{} delicately makes use of historical attacking information and conducts statistical test to avoid unnecessary model queries effectively. As a result,~\alagcro{} fools the target model effectively without incurring high query complexity. 

\noindent\textbf{Numerical  Benchmark.} We numerically demonstrate the effectiveness of \alagcro{} on the benchmark text classification experiments, across different model architectures and various datasets. \alagcro{} more adequately leverages each model query than state-of-the-art competitors. As a result, under unconstrained model query budget, \alagcro{} dramatically avoids 32.6\% queries in average and achieves competitive adversarial attacking performance. Furthermore, under constrained query budget, \alagcro{} outperforms others by attaining multi-times better attacking performance. 

\section{Related Work}
\subsection{Two Stage Attack In NLP}\label{sec.related_work.two_stage_nlp}

Most of the existing black-box word-based attackers consist of two main stages to generate adversarial examples. Specifically, the first stage calculates salience scores for each individual word based on various criteria. 
For example, \cite{deepwordbug} designs four scoring functions based model prediction,~\ie, {Replace-1}, {Temporal Head}, {Temporal Tail} and {Combined Score}. ~\cite{2019TextBugger}, ~\cite{ren-etal-2019-generating}, ~\cite{2020pwds}, ~\cite{2020-textfooler} and ~\cite{goodman2020fastwordbug} compute the importance of word as the confidence score deviation after deletion.
The tokens are then sorted via the importance score and fed into the second stage, wherein prescribed word transformation are employed on the selected tokens with significant impact onto the prediction. 
Both stages accompanies numerous model queries to compute either the salience score or the impact of each transformation.

\subsection{Query Efficiency in Black-Box NLP Attack}
The query-efficiency study in NLP black-box adversarial attack is quite limited. To the best of our knowledge, there are only a few related works. Two of them focus on optimizing the first-stage discussed in Section~\ref{sec.related_work.two_stage_nlp}. 
In particular, E2A~\cite{Explain2Attack} trains an interpretable agent model to learn word importance scores on substitute training corpus from similar fields of the target model.  \cite{hossam2021improved} follows the same procedure but improves the training process of E2A. Although both methods reduce the model query complexity, the performance of such approaches might be sensitive to the dataset selection and training quality. Furthermore, conducting such additional training requires lots of extra efforts, which is typically not convenient and scalable to large-scale intelligent NLP systems~\cite{brown2020language}. Transfer-based attackers can generate adversarial samples without accessing the target victim model~\cite{2021Model,wang2022transferable}, but have the same drawbacks as E2A. Recently,~\cite{berger2021don} generates adversarial samples within low model queries by random sampling, but such method fails on large-scale document set such as IMDB which limits its generality.

\subsection{Query Efficiency in Black-Box CV Attack}

In the field of computer vision black-box adversarial attack, the query efficiency has been developed to some extent. \cite{Bhagoji_2018_ECCV} partitions pixels into small groups and conduct group-wise queries. Then all pixels in one group operate the same perturbation based on the shared query. \cite{10.1145/3052973.3053009,2017-Delving,ma2021switching,247692,wang2020mgaattack} avoid querying target model by accessing substitute model instead, while the construction of the substitution is costly and time-consuming.
\cite{cheng2018query} reduces model queries by forming zeroth order optimization. ~\cite{li2020qeba} points out that the high query cost of black box attack is due to the high dimension of image input, and proposes an attack method based on dimension reduction. \cite{chen2020hopskipjumpattack} improves boundary attack by leveraging gradient estimation of decision boundary. Because of the discreteness of text space, these successful attack methods in CV are difficult to transfer to NLP area directly.

\section{Query Efficient Adversarial Attack}

\subsection{Target Problem Formulation}

In this section, we formulate the target problem of query-efficient adversarial attack for general natural language processing security tasks. To the best of our knowledge, such target problem is rarely well formulated across related computer vision and natural language processing applications.

For general NLP tasks, we aim at establishing an attacker $\mathcal{A}$ such that given a victim text set $\mathcal{X}$, for each text $X\in\mathcal{X}$,
the attacker $\mathcal{A}$ consumes the minimal number of model queries $\mathcal{Q}(\mathcal{A}, \mathcal{M}, X)$ to fool the model $\mathcal{M}$
Considering the popular classification problem, each document $X\in\mathcal{X}$ is associated with a ground truth label $Y_X$. The target problem becomes 
\begin{equation}\label{prob.main.original}\small
\begin{split}
\mathop {\text{minimize} }\limits_{\mathcal{A}}\ \sum_{X\in \mathcal{X}}\mathcal{Q}(\mathcal{A}, \mathcal{M}, X),
\text{s.t.}\ \mathcal{M}(\mathcal{A}(X))\neq Y_X\ \text{for}\ X\in \mathcal{X},
\end{split}
\end{equation}
where the adversarial document $\mathcal{A}(X)$ yields a wrongly predicted label by the model,~\ie,  $\mathcal{M}(\mathcal{A}(X))\neq Y_X$. However, problem~\eqref{prob.main.original} is typically intractable in practice since the constraint is usually hard to achieve for every document. As an alternative, we consider the following more relaxable unconstrained problem, 
\begin{equation}\label{prob.main.relaxable}\small
\begin{split}
\mathop {\text{minimize} }\limits_{\mathcal{A}}\ &\sum_{X\in \mathcal{X}}\mathcal{Q}(\mathcal{A}, \mathcal{M}, X) -\lambda r(\mathcal{A}, \mathcal{M}, X), 
\end{split}
\end{equation}
where $r(\mathcal{M}, \mathcal{A},X)$ is a regularization term~\cite{chen2022otov2,chen2021orthant,chen2021only} that measures the deviation of label between ground truth and adversarial example, \eg, $\norm{\mathcal{M}(\mathcal{A}(X))-Y_X}_2$, and $\lambda$ is some positive weighting parameter.
 
The majority of black-box adversarial attack in NLP tasks are two-stage attackers as discussed in Section~\ref{sec.related_work.two_stage_nlp},~\ie, the first-stage picks up important words and the second-stage proceeds various attack on these candidates. Both stages require numerous model queries. Consequently, the total number of queries equals to the summation of quantity from both stages.
In this paper, we concentrate on optimizing the cost of model query in the second stage,~\ie, $\mathcal{Q}_2(\mathcal{A}, \mathcal{M}, X)$ is free to be optimized. Therefore, the target problem becomes
\begin{equation}\label{prob.main.stagetwo}\small
\mathop {\text{minimize} }\limits_{\mathcal{A}}\ \sum_{X\in \mathcal{X}}\mathcal{Q}_2(\mathcal{A}, \mathcal{M}, X)-\lambda r(\mathcal{A}, \mathcal{M}, X).
\end{equation}

\subsection{Overview} 

To the best of our knowledge, we are the first to study optimizing the cost of model querying in the word transformation stage of black-box NLP attacks. To solve problem~\eqref{prob.main.stagetwo}, the proposed algorithm is stated in~Algorithm~\ref{alg:main.outline}.  In general, given a set of victim texts $\mathcal{X}$ and a target model $\mathcal{M}$, we iteratively attack each text $X$ in $\mathcal{X}$ and collect the intermediate information to form a table $\mathcal{T}$ as stated in Algorithm~\ref{alg:main.individual_attack}. Within each individual adversarial attack, we leverage the collected historical information $\mathcal{T}$ to refine the search space and avoid unnecessary model queries as Algorithm~\ref{alg:main.getcandidatelist}. As a result, we obtain the set of adversarial texts $\mathcal{X}_\text{adv}$ with much fewer number of queries than the existing methods. 
\begin{algorithm}[ht]
\small
	\caption{Query-Efficient Adversarial Attack}
	\label{alg:main.outline}
		\textbf{Input}: A set of victim texts $\mathcal{X}$ with the ground truth labels $\mathcal{Y}$,	target model $\mathcal{M}$, sentence similarity \verb|Sim(·)|, attack ratio $\epsilon$ and candidate list ratio $\gamma\in(0,1]$.\\
	    \textbf{Output}: A set of adversarial examples $\mathcal{X}_{\text{adv}}$.\\\vspace{-.15in}
	    \begin{algorithmic}[1]
	    \STATE Construct an empty table $\mathcal{\mathcal{T}}$ to store historical adversarial attack information. 
	    \FOR{each text $X$ in $\mathcal{X}$ and label $Y$ in $\mathcal{Y}$}
        \STATE{Get adversarial text $X_{\text{adv}}$ and update $\mathcal{T}$ via Algorithm~\ref{alg:main.individual_attack}.}\vspace{-.05in}
        $$
        X_{\text{adv}}\gets \text{Attack}(X, Y, \mathcal{M}, \mathcal{T}, \verb|Sim|, 
        \epsilon, 
        \gamma).
        $$\vspace{-.2in}
\STATE $\mathcal{X}_{\text{adv}}\gets \mathcal{X}_{\text{adv}}\bigcup X_{\text{adv}}$.
	    \ENDFOR
		\STATE \textbf{return } $\mathcal{X}_{\text{adv}}$.
	\end{algorithmic}
\end{algorithm}

Similarly to other state-of-the-arts black-box attackers, we design a two-stage algorithm for generating each individual adversarial text. In particular, the first stage picks up a set of words with high importance from the victim text $X$ based on various criteria as line~\ref{line.first_stage_start}-\ref{line.sortwordimportant} in Algorithm~\ref{alg:main.individual_attack}. Then in the second stage, the words are iterated by their ranked salience scores and transformed by some prescribed mechanisms to yield the ultimate adversarial example as line~\ref{line.second_stage_start}-\ref{line.second_stage_end} in Algorithm~\ref{alg:main.individual_attack}, wherein the transformations reply on the collected historical information $\mathcal{T}$ as shown Algorithm~\ref{alg:main.getcandidatelist}. All  replacement information is fed into the table $\mathcal{T}$ for further usage as line~\ref{line.get_soft_label}-\ref{line.update_table} in Algorithm~\ref{alg:main.individual_attack}.

\subsection{Word Importance Calculation}

In the stage one, we calculate the word salience score following the state-of-the-art ranker proposed in Textfooler~\cite{2020-textfooler}. In particular, suppose a victim text of $n$ words as $X\text{=\{}w_1, w_2,...,w_n\text{\}}$. To calculate the importance of $w_i$,~\ie, $I(w_i)$, we at first delete $w_i$ from $X$ to form  $X/\{w_{i}\}\text{=\{}w_1, ...,w_{i-1},w_{i+1},...,w_n\text{\}}$. Following Textfooler, the $I(w_i)$ has the following explicit forms. \\
$\sbullet[.75]$ Suppose the predicted label remains the same after deletion,~\ie, $\mathcal{M}(X)=\mathcal{M}\left(X/\{w_{i}\}\right)=Y_X$,
\begin{equation}
I(w_i)=\mathcal{M_{Y_X}}(X)-\mathcal{M_{Y_X}}\left(X/\{w_{i}\}\right).
\end{equation}
$\sbullet[.75]$ Suppose the predicted label is changed after deletion,~\ie, $\mathcal{M}(X)={Y_X}\neq {\hat{Y}_X}=\mathcal{M}(X/\{w_{i}\})$,
\begin{equation}
\begin{split}
I(w_i)=&\mathcal{M}_{Y_X}(X)-\mathcal{M}_{Y_X}\left(X/\{w_{i}\}\right)+\\
&\mathcal{M}_{\hat{Y}_X}(X)-\mathcal{M}_{\hat{Y}_X}\left(X/\{w_i\}\right),
\end{split}
\end{equation}
where $\mathcal{M}(X)$ represents the predicted label of $X$ by model $\mathcal{M}$, and $\mathcal{M}_{Y_X}(X)$ represents the predicted confidence score of $X$ by $\mathcal{M}$ on label $Y_X$.

\subsection{Query-Efficient Transformation Attack}

In the second stage, given the set of words with high importance $\mathcal{W}^*$ from the first stage,  we iterate the ranked words by their importance scores and perform a word replacement mechanism, wherein a proper word replacement should \textit{(i)} have similar semantic meaning with the original, and \textit{(ii)} confuse the target model to make wrong prediction. 

\noindent\textbf{Word Transformation.} Without loss of generality, consider the most popular synonym transformation as an illustrating example. For $w\in\mathcal{W}^*$, we at first establish the candidate set for replacement of $w$ $\mathcal{C}(w)$ from $\mathcal{S}(w)$ via Algorithm~\ref{alg:main.getcandidatelist}. 
Then as line~\ref{line.get_best_synonom_start}-\ref{line.replace_word_by_synonom} of Algorithm~\ref{alg:main.individual_attack}, during each replacement, we accomplish the replacement of $w$ as the candidate $c^*\in\mathcal{C}(w)$ achieving the largest deviation of prediction distribution,
\begin{equation}\label{prob.bestReplacement}
c^*:=\argmax_{c \in \mathcal{C}(w)}\ \mathcal{M}_{Y_X}(X_{\text{adv}})-\mathcal{M}_{Y_X}(X_{\text{adv}, w\to c })
\end{equation}
Consequently, the adversarial example $X_{\text{adv}}$ is constructed either after all words in $\mathcal{W}^*$ are exhausted and replaced by corresponding best synonym as~\eqref{prob.bestReplacement} or until the attack has been succeeded in advance as line~\ref{line.succeed_inadvance_start}-\ref{line.succeed_inadvance_end} of Algorithm~\ref{alg:main.individual_attack}.

The existing state-of-the-art methods solve problem~\eqref{prob.bestReplacement} by bruce-force exploiting the whole synonym space,~\ie, $\mathcal{C}(w)\equiv \mathcal{S}(w)$, so that calculating the prediction deviation requires $|\mathcal{S}(w)|$ model queries totally for each $w\in\mathcal{W}^*$. Thus the overall procedure performs  $\sum_{w\in\mathcal{W}^*}|\mathcal{S}(w)|$ model queries for the worst case, while there may exist numerous redundant queries to be optimized. 

\noindent\textbf{Pruning Transformation Space.} To increase the efficiency of word replacement, we prune the feasible space of transformation by identifying redundant word candidates to avoid potential unnecessary queries. Rather than utilizing the whole synonym set $\mathcal{S}(w)$, we establish the candidate set $\mathcal{C}(w)$ as a subset of $\mathcal{S}(w)$ which exhibits the largest prediction deviation in the history and has high confidence to be significant again verified via an efficient sorting algorithm and a statistical test.

To proceed, we at first initialize the candidate list of $w$ as $C_\text{initial}$ from the global table $\mathcal{T}$ associated with the previous attack information $\mathcal{H}(w,c)$ as line~\ref{line.get_candidate_list}-\ref{line.get_history_scores} in Algorithm~\ref{alg:main.getcandidatelist}, wherein $\mathcal{H}(w,c)$ represents the history of confidence score changes of replacing $w$ by $c$. Given a prescribed candidate list budget $\gamma \in (0, 1]$, we pick up a pivot candidate $c_{\text{pivot}}\in C_\text{initial}$ which has the $\lceil\gamma |\mathcal{C}_\text{initial}|\rceil$th largest confidence score deterioration in average as line~\ref{line.get_pivot_candidate} in Algorithm~\ref{alg:main.getcandidatelist}. The pivot candidate $c_{\text{pivot}}$ serves as the benchmark instance to filter out the candidates that are less likely to confuse the target model.

We then iterate all candidates from $\mathcal{C}_\text{initial}$. For each candidate $c$, a statistical test is employed to determine if $c$ can affect the model more significantly than the pivot candidate $c_\text{pivot}$ as line~\ref{line.statistical_test}. For simplicity, we represent the impact of replacement from $w$ to $c$ and $c_\text{pivot}$ as $\mu_c$ and $\mu_\text{pivot}$, respectively. Then, we propose to use the following one-sided test, 
\begin{equation*}
\text{null:} \mu_c \leq \mu_\text{pivot}\ \text{vs. alternative:}\ \mu_c > \mu_\text{pivot}.
\end{equation*}
To test these hypotheses, we make use of the historical $\mathcal{H}(w,c)$ and $\mathcal{H}(w,c_\text{pivot})$. Since these two populations to be compared may have unequal variance, Welch's two-sample one-tail t-test is performed here~\cite{welch1947generalization}. In particular, given the samples from $\mathcal{H}(w,c)$ and $\mathcal{H}(w,c_\text{pivot})$, we at first compute the t-statistic as 
\begin{equation}
t=\frac{\widebar{\mathcal{H}}(w,c)-\widebar{\mathcal{H}}(w,c_\text{pivot})}{\sqrt{{\hat{\sigma}_{c}^2}/{|\mathcal{H}(w,c)|}+{\hat{\sigma}_{c_\text{pivot}}^2}/{|\mathcal{H}(w,c_\text{pivot})|}}},
\end{equation}
where the $\bar{(\cdot)}$ and $\hat{\sigma}^2$ represent the sample mean and the variance estimator, respectively. Given a prescribed significance level $\alpha$, we then figure out the $(1-\alpha)$ quantile of the student's t-distribution with degree of freedom corresponding to that in the variance estimator, denoted as $t_{\alpha}^*$. Next, we reject the null hypothesis if $t\geq t_{\alpha}^*$.  If we fail to reject the null hypothesis that the impact of $c$ is less than that of $c_\text{pivot}$, then $c$ should be excluded from further consideration since replacing $w$ as $c$ largely generate negligible model confusion. Otherwise, we accept the alternative hypothesis of higher affect by $c$ and include it into the ultimate candidate list $\mathcal{C}(w)$ as line~\ref{line.hypythesis_rejected_start}-\ref{line.hypythesis_rejected_end}. 

\noindent\textbf{Complexity Analysis.} We here compare the complexity of model query by our proposed transformation attack. Since for each word $w\in\mathcal{W}^*$, we proceed a model-query-free statistical approach to filter out numerous negligible candidates. The remaining candidates after pruning quantize up to $\gamma|\mathcal{S}(w)|$, which result in an upper bound for the overall model query quantity as $\sum_{w\in\mathcal{W}^*}\gamma|\mathcal{S}(w)|$. Comparing to the $\sum_{w\in\mathcal{W}^*}|\mathcal{S}(w)|$  of the existing state-of-the-art methods, our method performs linearly faster in the manner of  model query in theory and will be further verified in Section~\ref{sec.experiments}.

\begin{algorithm}[ht]
\small
	\caption{Black Box Adversarial Attack }
	\label{alg:main.individual_attack}
	\textbf{Input:} Victim text $X$, ground truth label $Y$, target model $\mathcal{M}$, table $\mathcal{\mathcal{T}}$, similarity function $\text{Sim}(\cdot)$, and ratio $\epsilon, \gamma\in(0,1]$ for words and candidates selection.\\
    \textbf{Output:} Adversarial example $X_{\text{adv}}$. \\\vspace{-.15in}
	\begin{algorithmic}[1]
	    \STATE Initialize $X_{\text{adv}}\gets X$.
	    {\STATE Pre-process $X$ and get target word set $\mathcal{W}$. \label{line.crt.gettargetword}}
		\FOR{each word $w$ in $X$}\label{line.first_stage_start}
		    \STATE Compute the importance score $I(w)$. 
		\ENDFOR
		\STATE Set $\mathcal{W}^*$ by the top $\lfloor\epsilon|\mathcal{W}|\rfloor$ words from $\mathcal{W}$ via $I(w)$. \label{line.sortwordimportant}
		\FOR{each word $w$ in $W$}\label{line.second_stage_start}
			\STATE Get the  candidate list $\mathcal{C}_w$ from Algorithm~\ref{alg:main.getcandidatelist}.\label{line.update_table}\vspace{-.1in}
		    $$
		    \mathcal{C}_{w}\gets \text{CANDIDATE\_LIST}(w, \mathcal{T}, Y, \gamma).
		    $$\vspace{-.25in}
		    \STATE Get the recent soft label $\hat{y}_{\text{adv}}\gets \mathcal{M}_{Y}(X_{\text{adv}})$.\label{line.get_soft_label}
            \FOR{each candidate $c$ in $\mathcal{C}_w$}
		        \STATE $\hat{X}_c\gets \text{Replace}\ w\ \text{with}\ c\ \text{in}\ X_{\text{adv}}$.
		      \STATE Get hard/soft label $\mathcal{M}(\hat{X}_c)$ and $\hat{y}_c\gets\mathcal{M}_{Y}(\hat{X}_c)$. 
		      \STATE Update $\mathcal{T}[w,Y,c]\gets\mathcal{T}[w,Y,c]\bigcup (\hat{y}_{\text{adv}}-\hat{y}_c).$
		     \ENDFOR
		     \IF{there exists $c\in\mathcal{C}_w$ such that $\mathcal{M}(\hat{X}_c)\neq Y$}\label{line.succeed_inadvance_start}
		        \STATE Get the successful attack set $\mathcal{C}_{\text{success}}\subseteq \mathcal{C}_w$.
		        \STATE $c^*\gets \argmax_{c\in \mathcal{C}_{\text{success}}} \text{Sim}(X, \hat{X}_{c})$.
		        \STATE \textbf{Return } $X_{\text{adv}}\gets \hat{X}_{c^*}$.\label{line.succeed_inadvance_end}
		     \ELSE 
		        \STATE Computer the best candidate $c^*$ via~\eqref{prob.bestReplacement}.\label{line.get_best_synonom_start}
		        \STATE Replace $w$ with $c^*$ in $X_{\text{adv}}$.\label{line.replace_word_by_synonom}
		     \ENDIF\label{line.second_stage_end}
		\ENDFOR
		\STATE \textbf{Return} $X_{\text{adv}}$.
	\end{algorithmic}
\end{algorithm}

\begin{algorithm}[tb]
\small
	\caption{Compute Candidate List}
	\label{alg:main.getcandidatelist}
	\textbf{Input:} Victim word $w$, historical table $\mathcal{T}$, attack label $Y$, and budget parameter $\gamma$.\\
	\textbf{Output:} A  candidate list $\mathcal{C}({w})$.\\\vspace{-.15in}
	\begin{algorithmic}[1]
		\IF{$(w, Y)$ not found in $\mathcal{T}$}
	        \STATE Initialize the default candidate list $\mathcal{C}({w})$. 
	        \STATE \textbf{Return} $\mathcal{C}(w)$.
	    \ENDIF
	    \STATE Initialize $\mathcal{C}(w)\gets \emptyset$.
	    \STATE Get $\mathcal{C}_\text{initial}\gets \text{the word list of}\ \mathcal{T}(w,Y)$.\label{line.get_candidate_list}
	    \STATE Get history $\mathcal{H}(w, c)$ for each $c\in\mathcal{C}_\text{initial}$.  \label{line.get_history_scores}
	    \STATE Set $c_{\text{pivot}}$ as the pivot candidate with the $\lceil\gamma |\mathcal{C}|_\text{initial}\rceil$th largest average historical score.
	    \FOR{each $c \in \mathcal{C}_\text{initial}$}\label{line.get_pivot_candidate}
	        \STATE{Statistical test on $c_\text{pivot}$ impacts more than $c$  based on $\mathcal{H}(w, c)$ and $\mathcal{H}(w, c_\text{pivot})$. \label{line.statistical_test}}
	        
	        \IF{null hypothesis is rejected}\label{line.hypythesis_rejected_start}
	            \STATE Update $\mathcal{C}(w)\gets \mathcal{C}(w) \bigcup c$.
	        \ENDIF\label{line.hypythesis_rejected_end}
	    \ENDFOR
		\STATE \textbf{Return} $\mathcal{C}(w)$.
	\end{algorithmic}
\end{algorithm}

\section{Experiments}\label{sec.experiments}

\subsection{Datasets and Target Models}
To evaluate the effectiveness of our method, we apply \alagcro{} on the benchmark text classification task, which is perhaps the most popular and representative task studied in NLP adversarial attack area \cite{2020-textfooler}. 
In particular, the experiments cover various datasets such as \mr~\cite{mr_data},~\imdb~\cite{imdbdata}
and \yelp~\cite{yelp} as well as various model architectures ranging from \cnn~\cite{kim-2014-convolutional}, \lstm~\cite{6795963} to \bert~\cite{bert2019}. Remark here that though we demonstrate the effectiveness via text classification, our method is generic to be extended to other scenarios with minimal modifications.

\subsection{Baselines}

To quantitatively evaluate the performance, we compare \alagcro{} with two state-of-the-art adversarial attack methods PWWS~\cite{ren-etal-2019-generating} and Textfooler (TF)~\cite{2020-textfooler} employed on the text classification tasks. We include random.CS~\cite{berger2021don} on~\imdb{} to demonstrate the generality of~\alagcro{} on reducing query complexity for various text datasets.
We exclude \cite{Explain2Attack,hossam2021improved}, since their strategies focus on the first stage and are complementary to our methodology.  

\subsection{Experimental Settings}
In our black-box experiments, the attacker can only access the input text and the output prediction (confidence scores) by querying the victim model. 
The default candidate list in Algorithm~\ref{alg:main.getcandidatelist} is set as the synonym set for each word, which is constructed similarly to TF. Specifically, given the word embeddings~\cite{mrksic-etal-2016-counter}, we compute the cosine similarity across different instances, then pick up the ones with top $N$ sufficiently high similarity (greater than $\delta$) as the synonyms of each word. As other literatures, we empirically set $N=50$ and $\delta=0.5$~\cite{2020-textfooler}.
The budget hyper-parameter $\gamma$, which controls the number of candidate words, and the significance level are set to be $0.3$ by default.

\subsection{Automatic Evaluation Metrics}
We evaluate the performance by four popular metrics.  
Adv Accuracy---the accuracy of target model on the adversarial examples.
Query Num---the number of model queries during adversarial attack.
Perturbation Rate---the rate of tokens substituted in the origin texts. 
Semantic similarity---the semantic similarity between the adversarial examples and the victim examples measured by the popular open-source tool~\cite{bertencode}.
The first two measure the effectiveness of adversarial attack, and the later ones measure the attack invisibility.

\subsection{Automatic Numerical Evaluation}
\label{sec.results}
The evaluation results of~\alagcro{} and strong baselines are reported in Table~\ref{table.attack} and~\ref{table.random.CS}, wherein the best numbers are marked as bold for the ease of comparison. In general, our method achieves efficient and effective adversarial attacks for all experiments with detailed descriptions as follows. 

\begin{figure}[]
\centering
\subfigure[\cnn]
{
    \begin{minipage}[b]{.3\linewidth}
        \centering
        \includegraphics[scale=0.124]{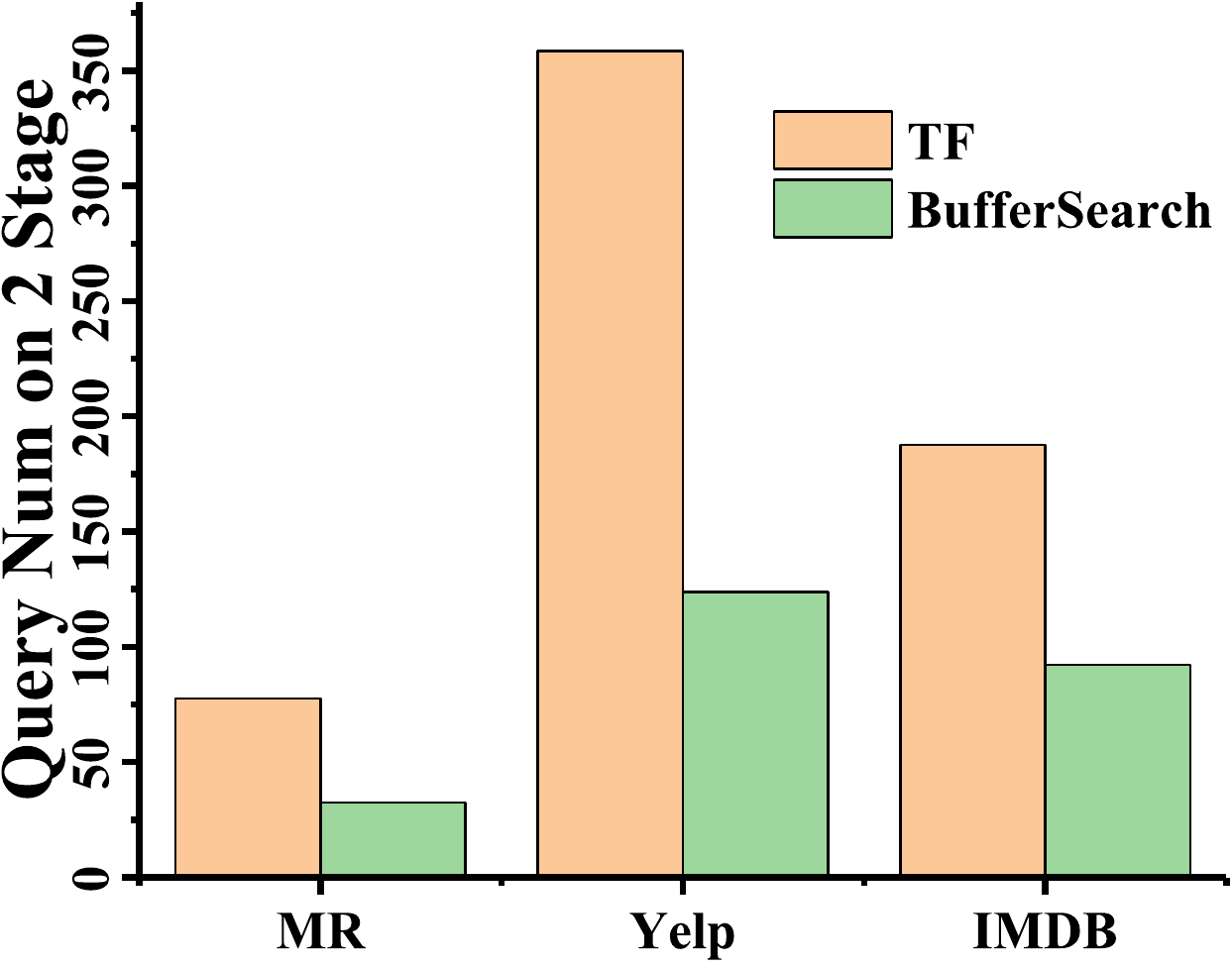} 
    \end{minipage}
    \label{figure.cnn_query}
}
\subfigure[\lstm]
{
    \begin{minipage}[b]{.3\linewidth}
        \centering
        \includegraphics[scale=0.124]{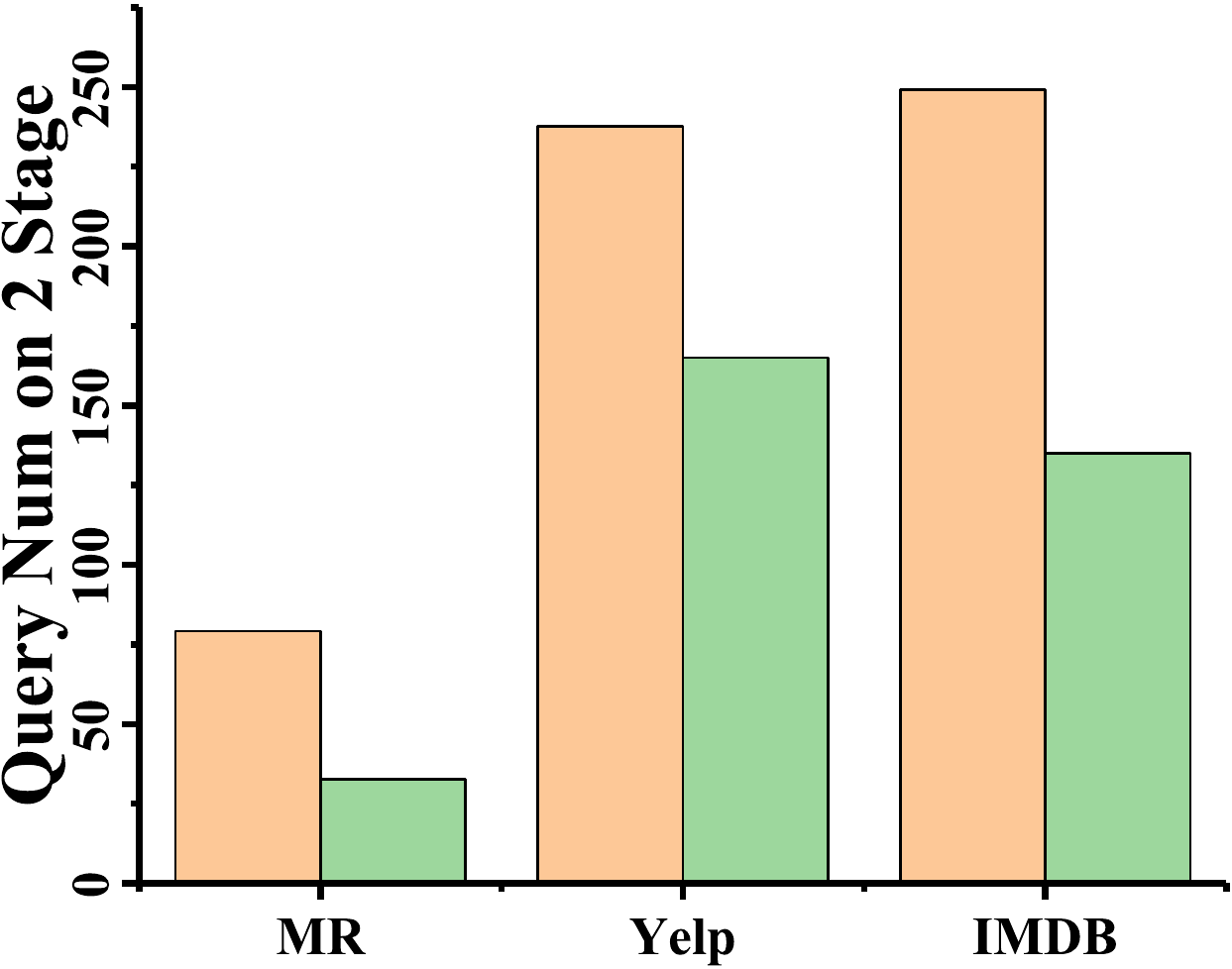} 
    \end{minipage}
    \label{figure.lstm_query}
}
\subfigure[\bert]
{
    \begin{minipage}[b]{.3\linewidth}
        \centering
        \includegraphics[scale=0.124]{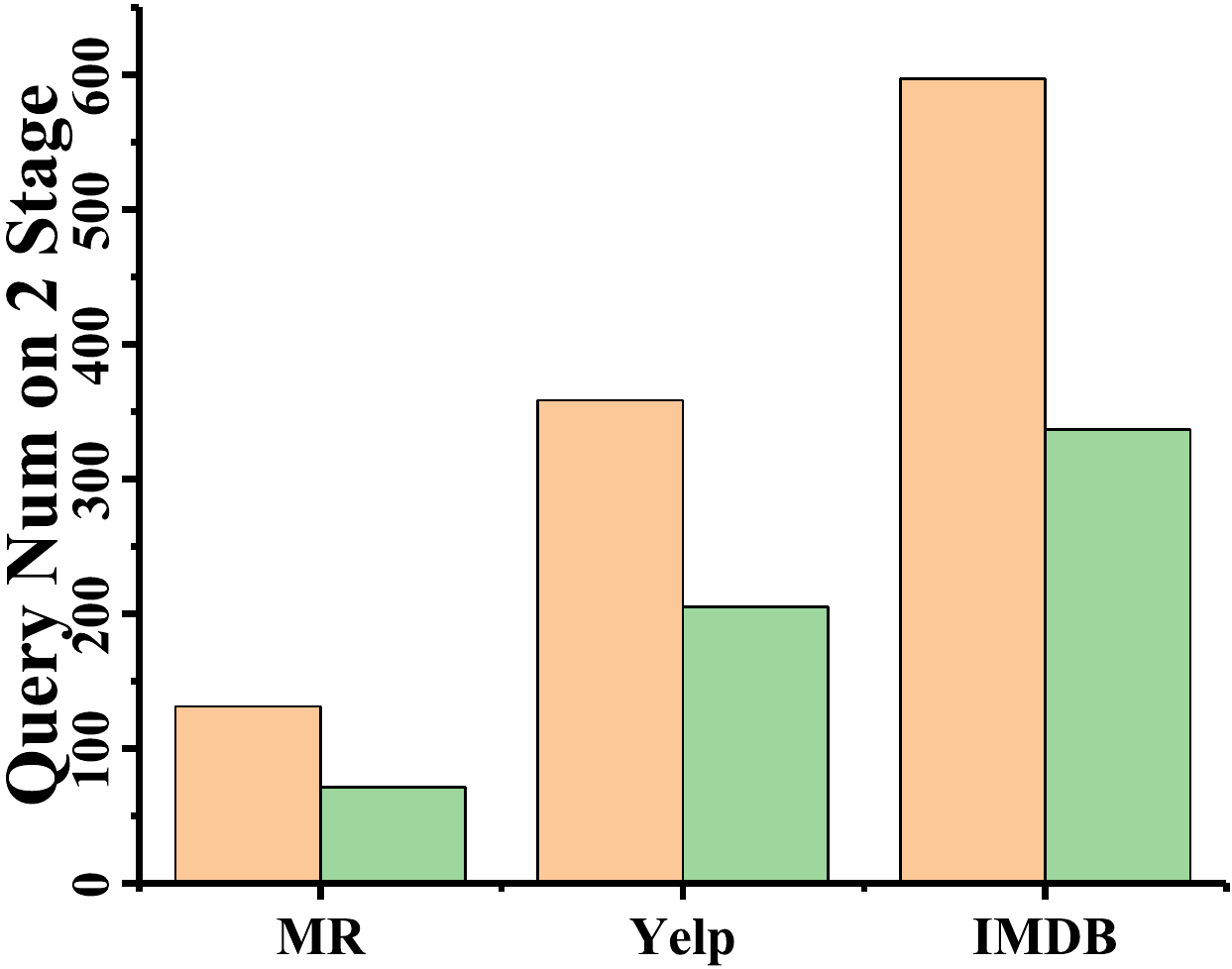}
    \end{minipage}
    \label{figure.bert_query}
}
\vspace{-0.1in}
\caption{Query cost comparison between TF and \alagcro{} over \cnn, \lstm{} and \bert{} on \mr, \yelp{} and \imdb.} \label{figure.second_query}
\vspace{-0.2in}
\end{figure}

\noindent
\textbf{Efficiency.}  Compared to the state-of-the-art baselines,~\ie, TF and PWWS, \alagcro{} requires significantly fewer model queries during the whole attacking procedure. In particular, \alagcro{} reduces the query cost  by 32.6\% in average across all experiments compared to TF, where a 42.2\% reduction is even achieved in the WordCNN on MR experiment. The superiority of \alagcro{} on query efficiency is due to the leverage of historical information to avoid unnecessary queries in the second stage as shown in Figure~\ref{figure.second_query}. PWWS is not comparable to \alagcro{} and TF in terms of query efficiency, which is  $6-10$ times costly than our method. As drawn in Table~\ref{table.random.CS}, random.CS fails to effectively attack IMDB(BERT) though achieves the lowest query cost.

\noindent
\textbf{Effectiveness.} Although \alagcro{} dramatically reduces the model queries, it still achieve outstanding attacking performance. Our method can greatly fool all the three target models. Even for BERT with the best robustness, the accuracy can be reduced from about 90\% to no more than 12\% on all benchmark datasets. In fact,~\alagcro{} achieves the best (lowest) and the second best After-Attack Accuracy on two and six out of nine experiments, respectively, and competitive accuracy on the remaining one test with the gap less than $1.1\%$. All methods perform closely in the manner of perturbation rate. For example, \alagcro{} is as low as 2.9\% on the IMDB under WordCNN. That means for a input containing 100 tokens,~\alagcro{} only needs to replace three words or less to generate an adversarial output. Similar observations can be found in the semantic similarity, which is inversely proportional to the disturbance rate.

\noindent
\textbf{Discussion.} Throughout the experiments, it is no doubt that~\alagcro{} is the best in terms of query efficiency and attacking performance comprehensively. The results demonstrate the effectiveness of our proposed algorithm. On the other hand, although TF and PWWS consume dramatically more resources to query the victim models, the gains they received on the attacking performance are negligible. It consequently reveals the existence of numerous query redundancy. We hope that \alagcro{} establishes a baseline for future studies to further push the boundary.

\subsection{Human Evaluation}

To further verify the quality of generated adversarial examples by~\alagcro{}, we proceed additional human evaluation following the settings of~\cite{2020-textfooler}. In particular, we designed three experiments in laboratory environment to evaluate the following aspects. (i) \textit{Grammatical naturalness score.} We asked human judges to put a score (1-5) to evaluate the grammatical validity of a sentence. (ii) \textit{Label prediction.} We showed human judges $50$ samples randomly selected from MR training data, then asked them to classify the motion adversarial example as either negative or positive. (iii) \textit{Semantic similarity.} We asked judges to put a score $0, 0.5, 1$ to evaluate the semantic similarity between two sentences.

\begin{figure}[t]
    \centering
    \includegraphics{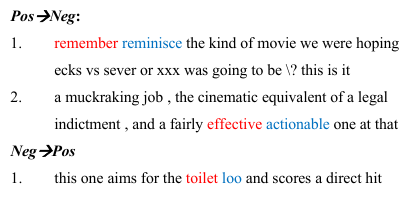}
    \vspace{-0.1in}
    \caption{Adversarial samples generated by \alagcro{} from~\mr{} (\bert) dataset. Target and replacement words are resp. in {\color{red}red} and {\color{blue}blue}. The label predicted by~\bert{} is changed.}
    \label{figure.case}
    \vspace{-0.1in}
\end{figure}

To proceed, we randomly selected $100$ adversarial samples generated on dataset~\mr{} under~\bert{} and their corresponding origin samples. Each experiment was evaluated by 5 human judges, which average results are shown in Table~\ref{table.human}. As examples shown in Figure~\ref{figure.case}, the adversarial samples generated by~\alagcro{} maintain the semantic information of origin input and successfully fool the victim target model.

\begin{table}[h]

\centering
\small
\begin{tabular}{lcc}
\hline
Text & origin text & adv text \\ \hline
Prediction accuracy &72.3  &65.2  \\
Grammatical naturalness &3.56  &3.42  \\
Semantic similarity & \multicolumn{2}{c}{0.869} \\ \hline
\end{tabular}
\caption{Average score of human evaluation.} \label{table.human}
\vspace{-.25in}
\end{table}

\begin{table*}[h]
\scriptsize \centering
\begin{tabular}{lccccccccc}
\hline
\multicolumn{10}{c}{\cnn} \\ \hline
\multicolumn{1}{l}{} & \multicolumn{3}{c}{\mr} & \multicolumn{3}{c}{\yelp} & \multicolumn{3}{c}{\imdb} \\
 & TF & PWWS & \alagcro & TF & PWWS & \alagcro & TF & PWWS & \alagcro \\
Original Accuracy & 78.0 & 78.0 & 78.0 & 94.0 & 94.0 & 94.0 & 89.2 & 89.2 & 89.2 \\
Adv Accuracy & \textbf{1.2} & 1.4 & 1.5 & \textbf{0.1} & 1.1 & 1.2 & \textbf{0} & 0 & \textbf{0} \\
\% Perturbed & 12.6 & \textbf{10.9} & 12.7 & 6.4 & \textbf{5.7} & 7.7 & 2.5 & \textbf{2.2} & 2.9 \\
Semantic Similarity & \textbf{0.963} & 0.960 & 0.957 & \textbf{0.989} & \textbf{0.989} & 0.987 & \textbf{0.998} & \textbf{0.998} & 0.997 \\
Query Num & 97.7 & 285.1 & \textbf{52.3} & 384.3 & 1855.7 & \textbf{270.6} & 441 & 3302.7 & \textbf{345.5} \\
\hline
\multicolumn{10}{c}{\lstm} \\ \hline
Original Accuracy & 80.7 & 80.7 & 80.7 & 96.0 & 96.0 & 96.0 &89.8 & 89.8 & 89.8 \\
Adv Accuracy & \textbf{1.1} & 2.1 & 1.2 & \textbf{0.4} & 2.1 & 1.7 & \textbf{0.3} & 0.6 & \textbf{0.3} \\
\% Perturbed & 12.9 & \textbf{11.2} & 13.1 & \textbf{7.2} & 8.1 & 9 & \textbf{3.1} & \textbf{3.1} & 3.7 \\
Semantic Similarity & \textbf{0.960} & \textbf{0.960} & 0.955 & \textbf{0.987} & 0.984 & 0.983 & \textbf{0.997} & 0.996 & 0.995 \\
Query Num & 99.1 & 285.9 & \textbf{52.3} & 429.8 & 1863.6 & \textbf{313.5} & 497.0 & 3306.6 & \textbf{385.1} \\
\hline
\multicolumn{10}{c}{BERT} \\ \hline
Original Accuracy & 90.4 & 90.4 & 90.4 & 97.0 & 97.0 & 97.0 & 90.9 & 90.9 & 90.9 \\
Adv Accuracy & \textbf{9.5} & 18.5 & 10.5 & \textbf{0.6} & 7.8 & 1.9 & \textbf{11.2} & 13.5 & 11.7\\
\% Perturbed & 17.7 & \textbf{13.6} & 18.1 & \textbf{9.4} & 9.5 & 10.3 & \textbf{3.8} & 5.0 & 4.2 \\
Semantic Similarity & 0.939 & \textbf{0.946} & 0.933 & \textbf{0.981} & 0.978 & 0.979 & \textbf{0.995} & 0.992 & 0.994 \\
Query Num & 144.7 & 285.2 & \textbf{86.3} & 502.1 & 1811.7 & \textbf{347.4} & 693.8 & 3240.5 & \textbf{501.6} \\
\hline
\end{tabular}
\caption{Results under unlimited query budget. (Original Accuracy:the model prediction accuracy on $1000$ origin samples, Adv Accuracy:the model accuracy after attack, \% Perturbed:the percentage of substitution of the original text, Semantic Similarity:the semantic similarity between original and adversarial samples, Query Num:the cost of query in attack, Average Text Length:the average length of original text.)}
\label{table.attack}
\vspace{-0.1in}
\end{table*}

\begin{table}[]
\small
\begin{tabular*}{\linewidth}{@{}@{\extracolsep{\fill}}lcccc@{}}\hline
&Ori Acc & Adv Acc & \% Perturbed & Query Num \\ \hline
random.CS&90.9              & 62.8         & 10.7      & \textbf{125.1} \\
\alagcro&90.0 & \textbf{11.7}&\textbf{4.2} &501.6 \\ \hline 
\end{tabular*}
\caption{\small \alagcro{} versus random.CS on IMDB(BERT).}
\label{table.random.CS}
\vspace{-0.1in}
\end{table}

\section{Discussion}

\subsection{Limited Query Budget Setting}
\begin{figure}[]
\centering
\subfigure[MR]
{
    \begin{minipage}[b]{.3\linewidth}
        \centering
        \includegraphics[scale=0.125]{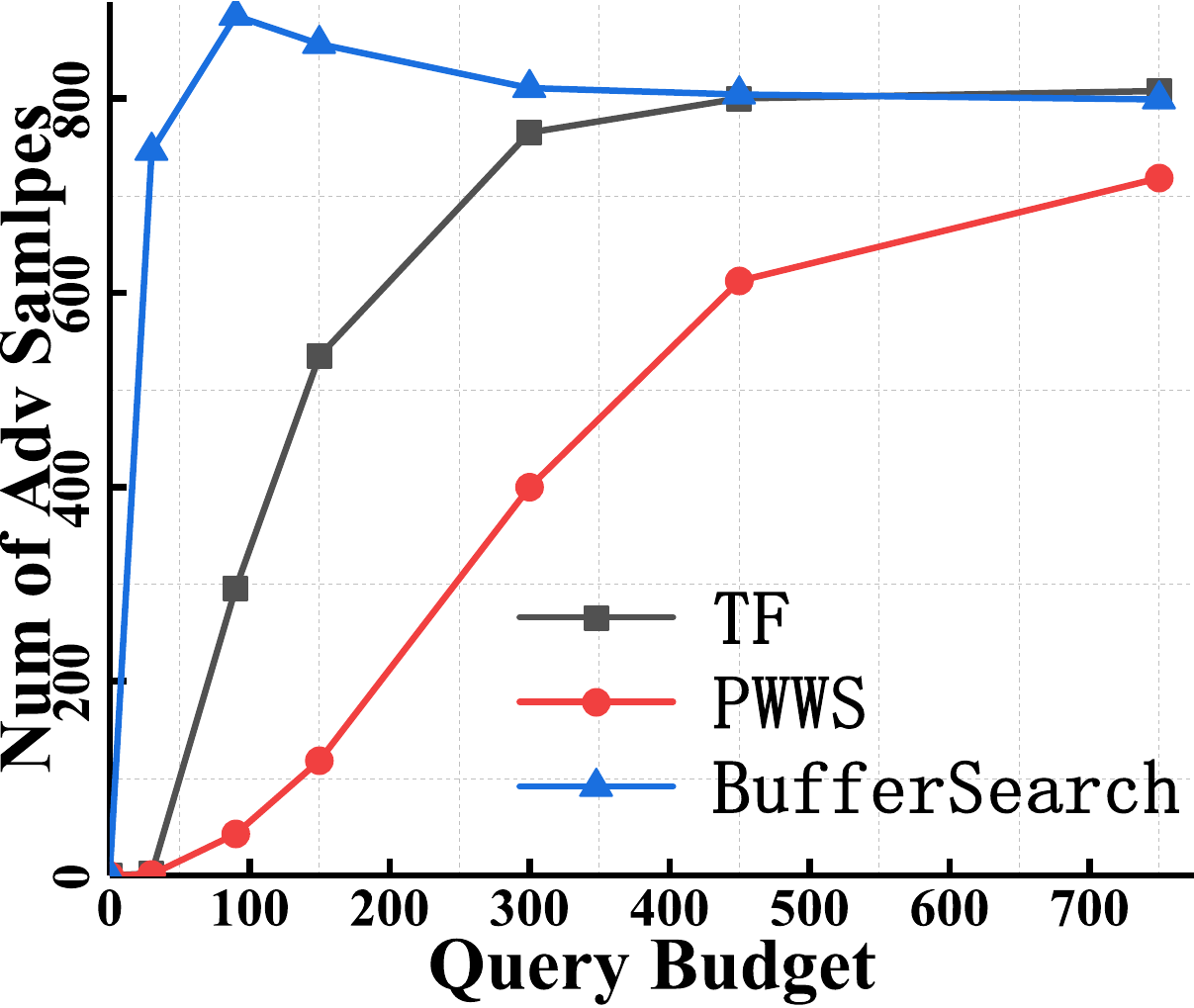}
    \end{minipage}
    \label{figure.mr_budget}
}
\subfigure[Yelp]
{
    \begin{minipage}[b]{.3\linewidth}
        \centering
        \includegraphics[scale=0.125]{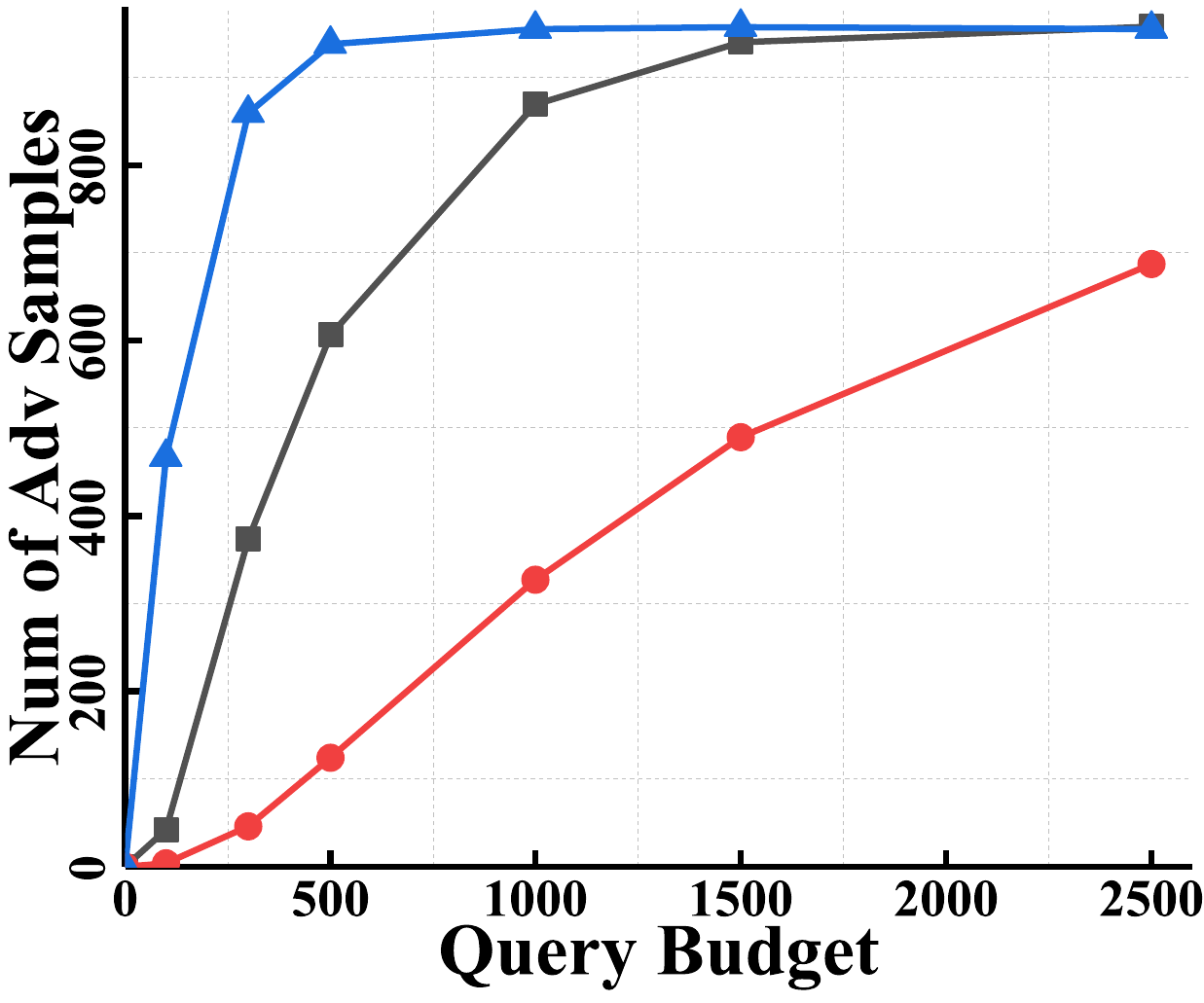}
    \end{minipage}
    \label{figure.yelp_budget}
}
\subfigure[IMDB]
{
    \begin{minipage}[b]{.3\linewidth}
        \centering
        \includegraphics[scale=0.125]{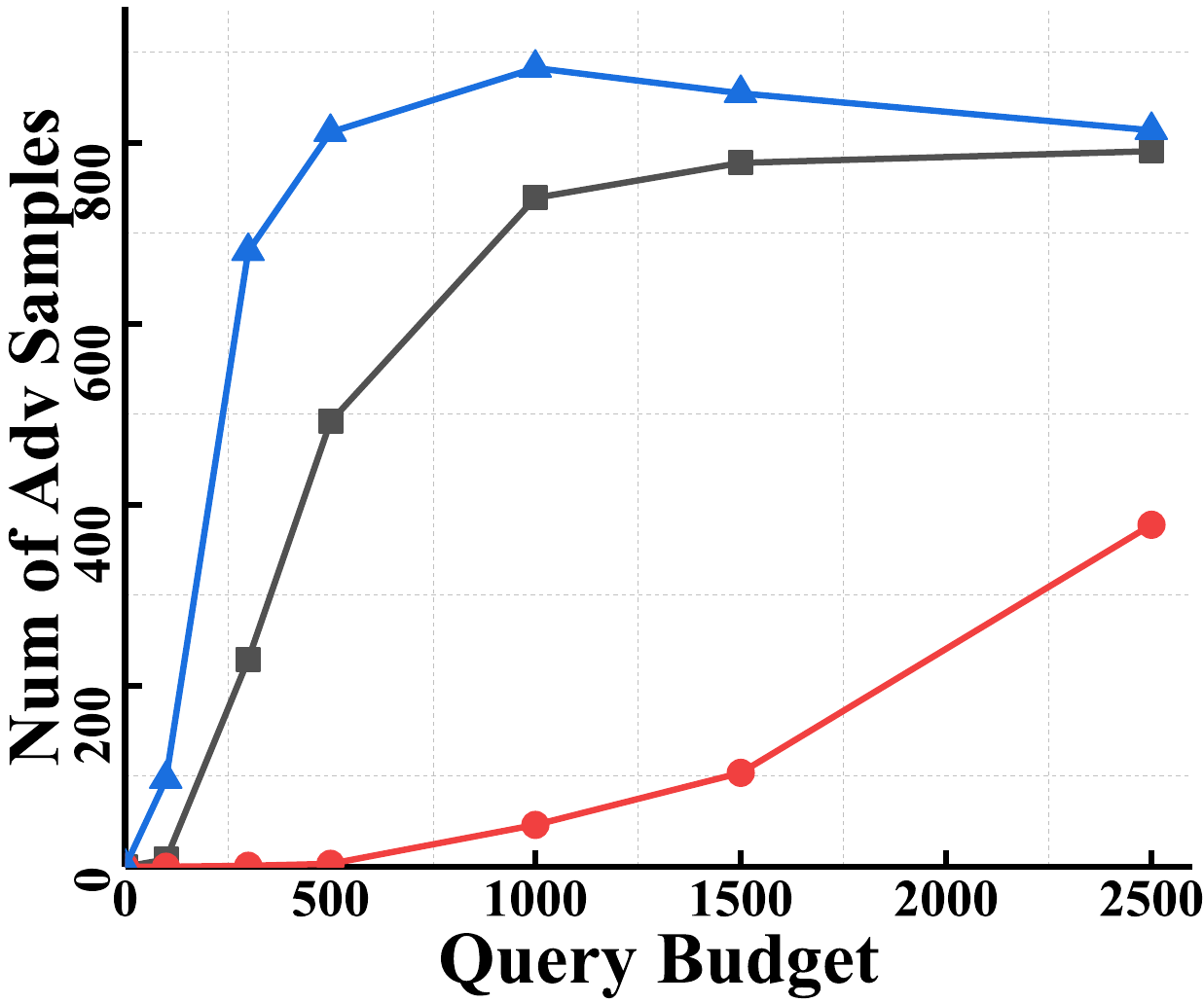}
    \end{minipage}
    \label{figure.imdb_budget}
}
\caption{\vspace{-.02in}The number of adversarial samples generated under varying the query budget $Q_\text{max}$ on \bert{}.
}\label{figure.budget}
\vspace{-0.1in}
\end{figure}

Besides the setting of unlimited query budget in Section~\ref{sec.results}, we study the performance of the attackers  under restricted query budget $Q_\text{max}$'s. This constraint forces the attacker to return adversarial examples within $Q_\text{max}$ queries. We test a variety of $Q_\text{max}$'s from $\{30,90,150,300,450,750\}$ for~\mr{}, and $\{100,300,500,1000,1500,2500\}$ for~\imdb{} and~\yelp{}. The performance is measured by the number of adversarial examples that successfully fool the victim models. As displayed on Figure~\ref{figure.budget} 
, our method performs the best across all models and datasets. In particular, ~\alagcro{} performs multiple times better than TF and PWWS under low $Q_\text{max}$'s. Then the successful attacking rates of all methods converge as the query budget increases. These experiments serve as a strong evidence that~\alagcro{} explored the word transformation space more effectively than other competitors.

\subsection{Different Size of Historical Information}
\begin{figure}[]\centering
\subfigure[\imdb]
{
    \begin{minipage}[b]{.4\linewidth}
        \includegraphics[scale=0.145]{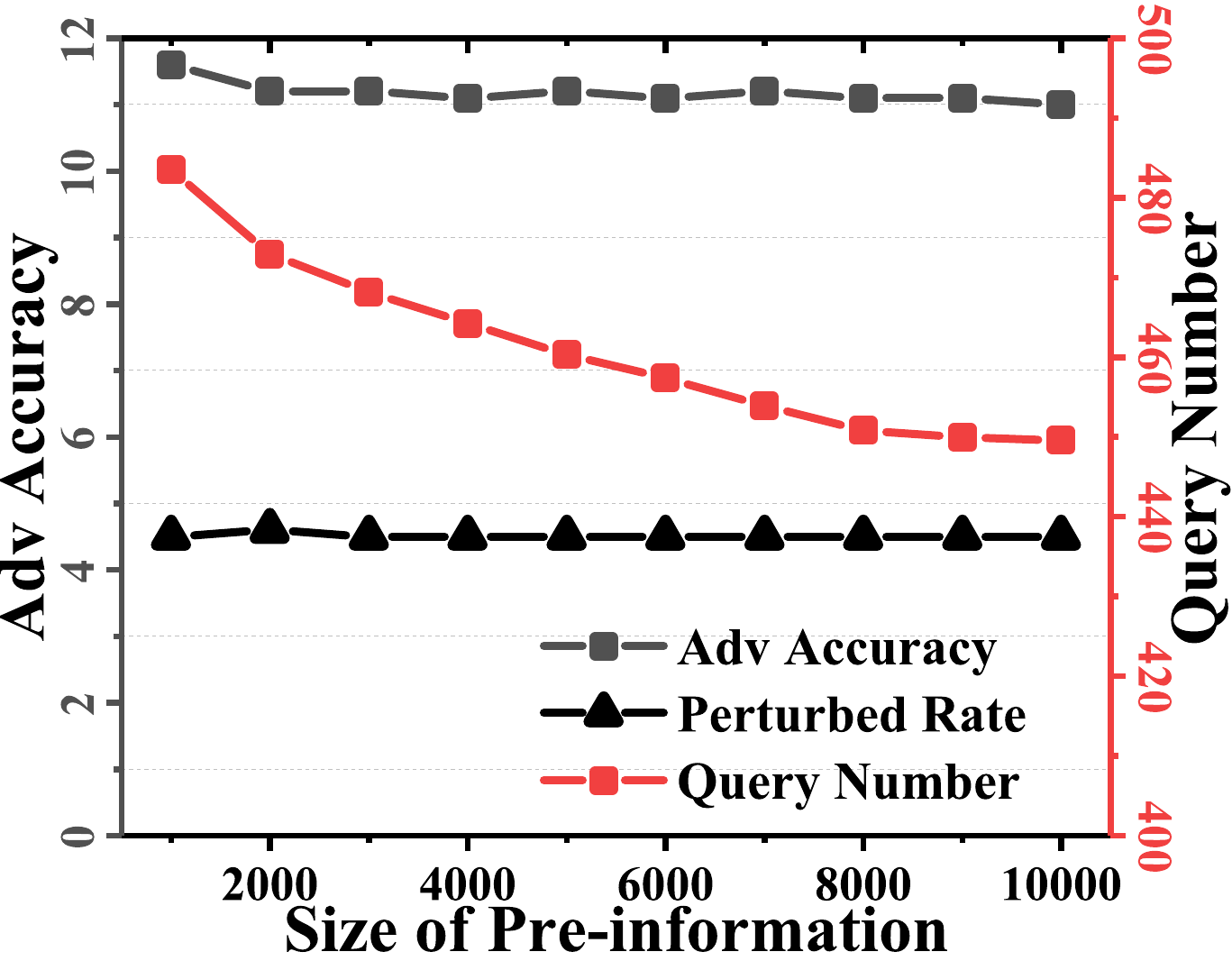}
    \end{minipage}
    \label{figure.history_imdb}
}
\subfigure[\yelp]
{
    \begin{minipage}[b]{.4\linewidth}\centering
        \includegraphics[scale=0.145]{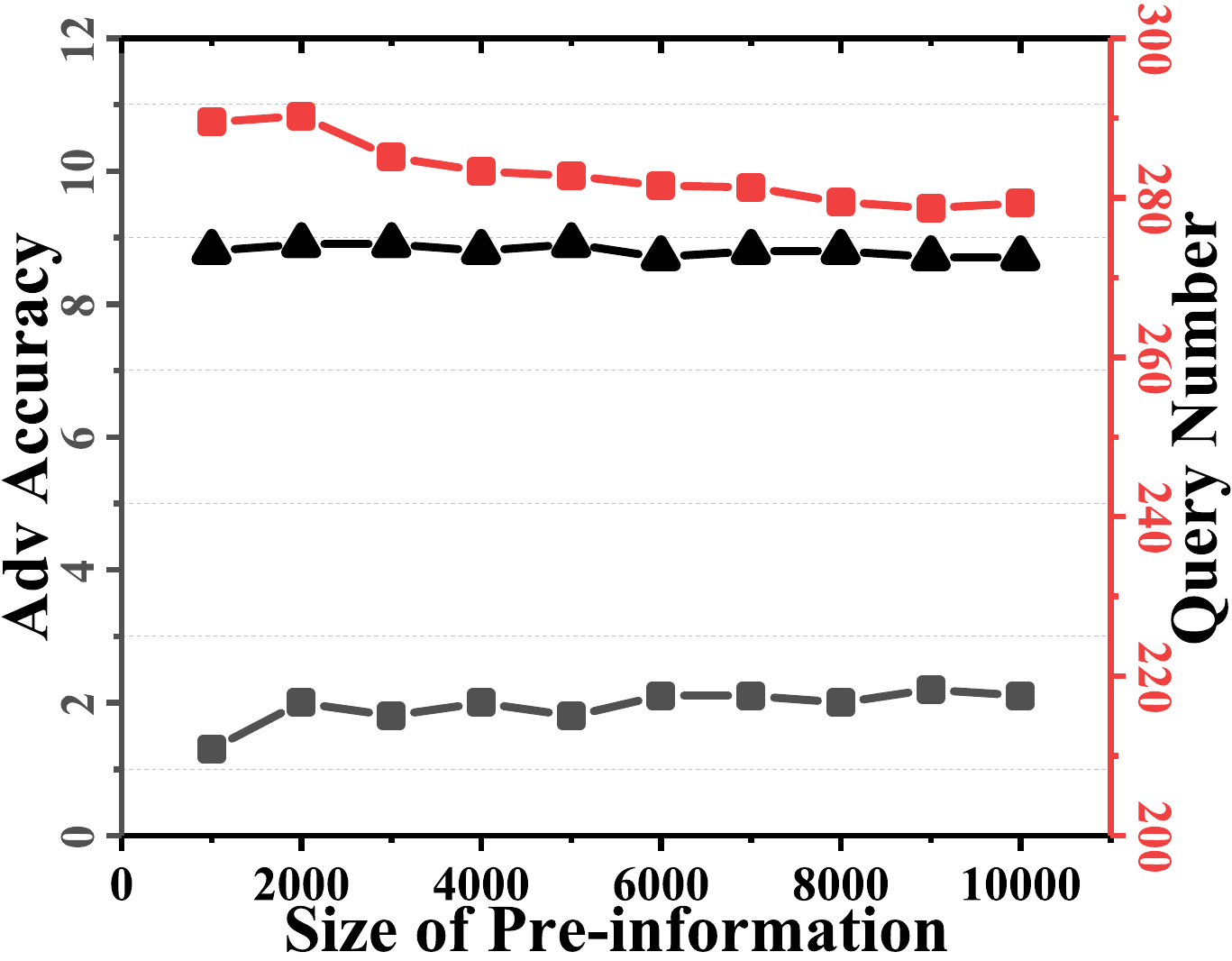} 
    \end{minipage}
    \label{figure.history_yelp}
}
\vspace{-0.1in}
\caption{Attacking performance based on varying size of historical information IMDB (BERT(a)) and Yelp (WordLSTM(b)).}\label{figure.history}
\vspace{-0.1in}
\end{figure}

To explore the impact of different size of historical information onto the attacking performance of~\alagcro{}, we randomly select ten subsets with size ranging from $10^3$ to $10^4$ from Yelp and~\imdb{} test datasets, then employ our method onto these subsets to collect historical tables. These tables are then leveraged into attacking the original $10^3$ test samples for evaluation. As drawn in Figure~\ref{figure.history}, there is slight but not significant improvement when more historical information is leveraged. It reveals that \alagcro{} does not highly rely on the historical accumulation after information saturation.

\begin{table}[tbh]
\tiny
\begin{tabular*}{\linewidth}{@{}@{\extracolsep{\fill}}llccc@{}}
\hline
 &  & \cnn & \lstm & \bert \\ \hline
\multirow{3}{*}{\mr} & \cnn & 1.5 & 54.0 & 85.7 \\
 & \lstm & 51.3 & 1.2 & 84.5 \\
 & \bert & 57.7 & 59.6 & 10.5 \\ \hline
\multirow{3}{*}{\yelp} & \cnn & 1.2 & 83.2 & 91.3 \\
 & \lstm & 80.3 & 1.7 & 90.4 \\
 & \bert & 85.0 & 85.1 & 1.9 \\ \hline
\multirow{3}{*}{\imdb} & \cnn & 0.0 & 76.0 & 84.2 \\
 & \lstm & 81.0 & 0.3 & 81.7 \\
 & \bert & 70.5 & 83.3 & 11.7 \\ \hline
\end{tabular*}
\caption{Transferability of adversarial examples generated by row-wise models and datasets then evaluated on column-wise models. 
}
\label{table.transfer}
\end{table}

\subsection{Transferability}
The transferability of adversarial examples refers to the property that the same input can successfully confuse different models~\cite{srivastava2014dropout}. We evaluate the transferability on $3$ datasets across \bert{}, \cnn{} and \lstm{}. As shown in Table~\ref{table.transfer}, the adversarial samples generated by attacking one model fool other models to varying extents. \bert{} exhibits the best robustness to transferable attacks.

\subsection{Adversarial Training}
In order to verify the effect of adversarial examples on model robustness, we retrain the \bert{} on MR over the training set augmented by the adversarial data generated by \alagcro{}. We then employ \alagcro{} on the retrained \bert{} and report the results in Table \ref{table.adv_train}. It is apparent that the model becomes more robust to the adversarial attacks after adversarial training, since the after attack accuracy, perturbation rate and query number significantly increased.

\begin{table}[tbp]
\centering
\small
\resizebox{\linewidth}{!}{
\begin{tabular}{ccccc}
\hline
\multirow{2}{*}{Adv Training} & \multicolumn{2}{c}{Accuracy (\%)} & \multirow{2}{*}{\# of queries} 
 & \multirow{2}{*}{Perturbation} \\ 
 \cline{2-3}
 & Original & Adv & & \\
 \hline
\xmark  & 87.9 & 3.5 & 63.0 & 14.7 \\ 
\checkmark & 85.9 & 14.1 & 81.6 & 16.1 \\ \hline
\end{tabular}
}
\caption{Attacking~\bert{} with(out) adversarial training on~\mr{}.}
\label{table.adv_train}
\vspace{-0.2in}
\end{table}

\section{Conclusion And Future Work}
We propose a query efficient attack method \alagcro{} on text classification task in black-box setting. Extensive experiments show that \alagcro{} can generate high quality adversarial samples with significantly fewer queries. Our work establishes baselines for the future query efficient attack studies in NLP related tasks.

\bibliographystyle{named}
\bibliography{ijcai22}

\appendix

\end{document}


\maketitle

\appendix


\begin{figure*}[h]
\centering
\subfigure[\mr{}]
{
    \begin{minipage}[b]{.3\linewidth}
        \centering
        \includegraphics[scale=0.245]{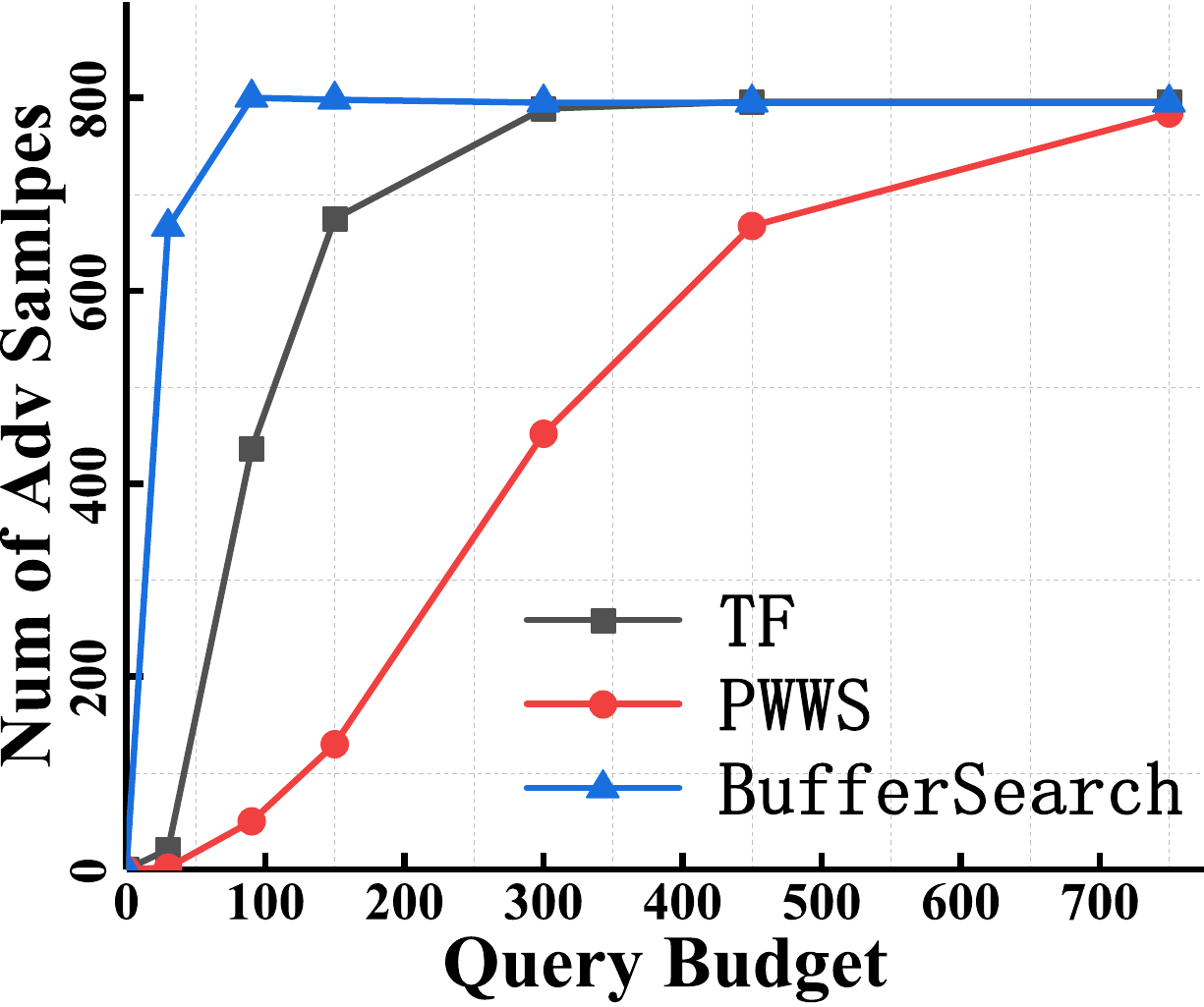}
    \end{minipage}
    \label{figure.mr_budget_lstm}
}
\subfigure[\yelp{}]
{
    \begin{minipage}[b]{.3\linewidth}
        \centering
        \includegraphics[scale=0.245]{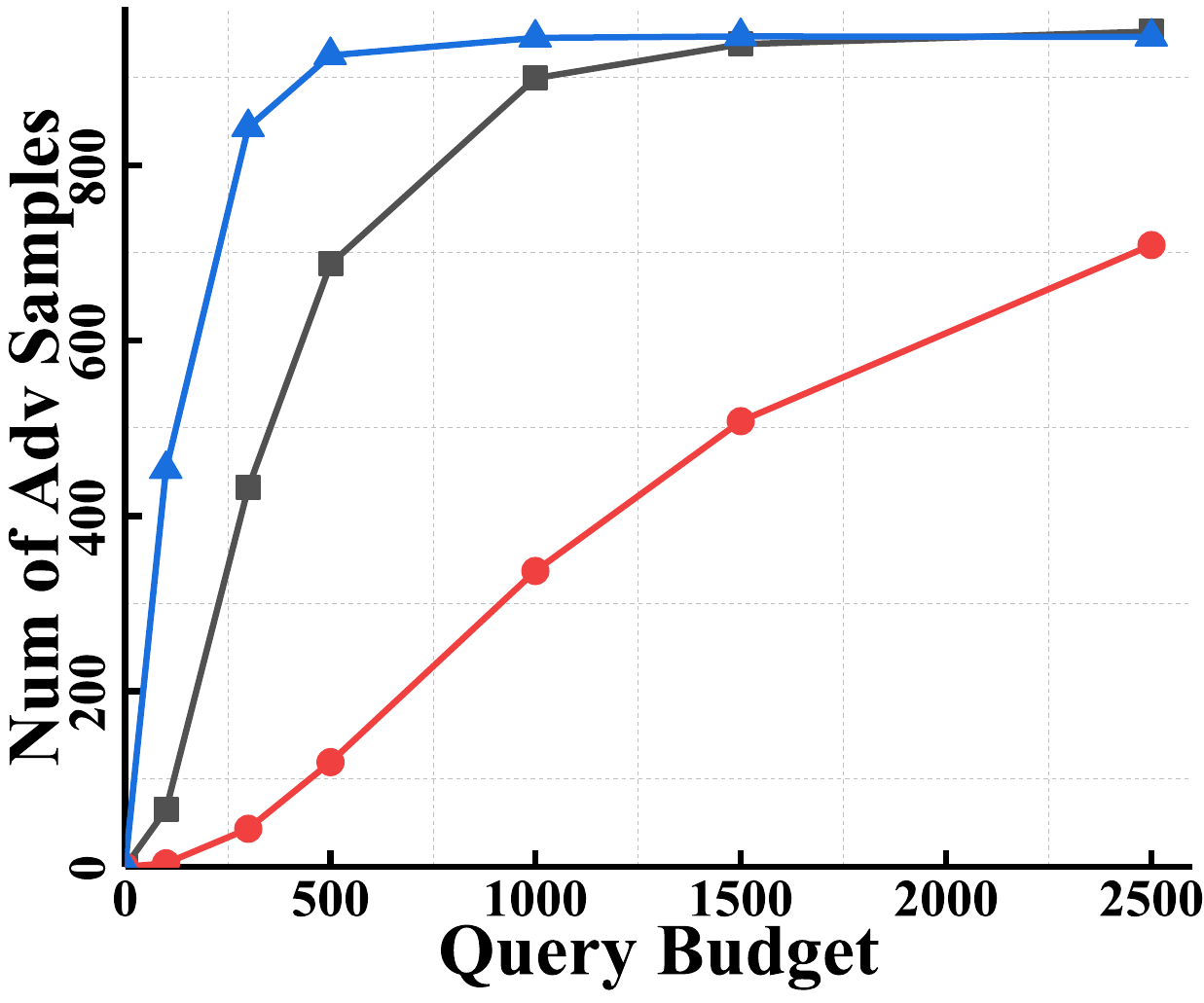}
    \end{minipage}
    \label{figure.yelp_budget_lstm}
}
\subfigure[\imdb{}]
{
    \begin{minipage}[b]{.3\linewidth}
        \centering
        \includegraphics[scale=0.245]{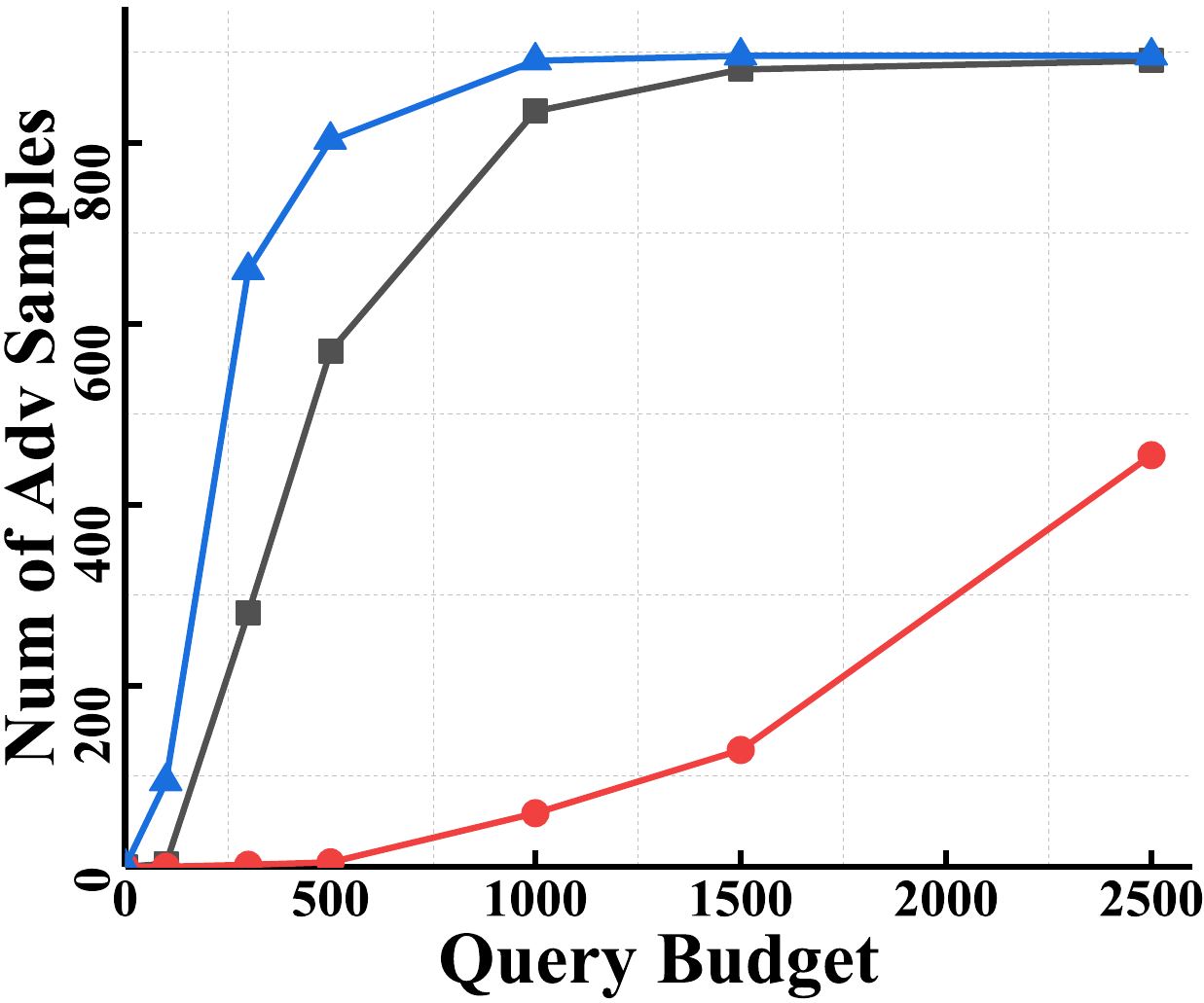}
    \end{minipage}
    \label{figure.imdb_budget_lstm}
}
\caption{The number of successful adversarial samples generated under varying the query budget $Q_\text{max}$ on WordLSTM.}
\label{figure.budget_lstm}
\end{figure*}

\begin{figure*}[h]
\centering
\subfigure[\mr{}]
{
    \begin{minipage}[b]{.3\linewidth}
        \centering
        \includegraphics[scale=0.245]{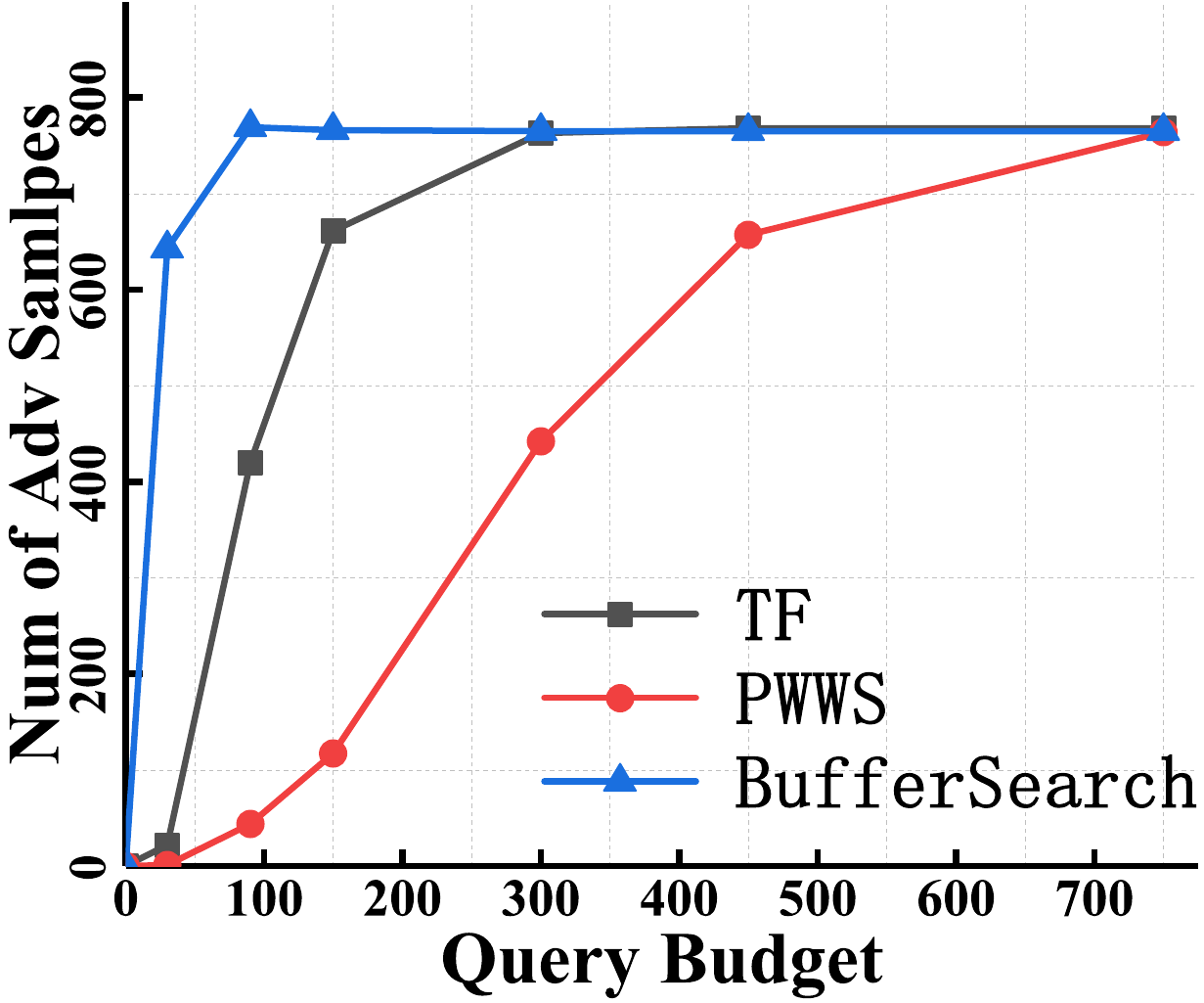}
    \end{minipage}
    \label{figure.mr_budget_cnn}
}
\subfigure[\yelp{}]
{
    \begin{minipage}[b]{.3\linewidth}
        \centering
        \includegraphics[scale=0.245]{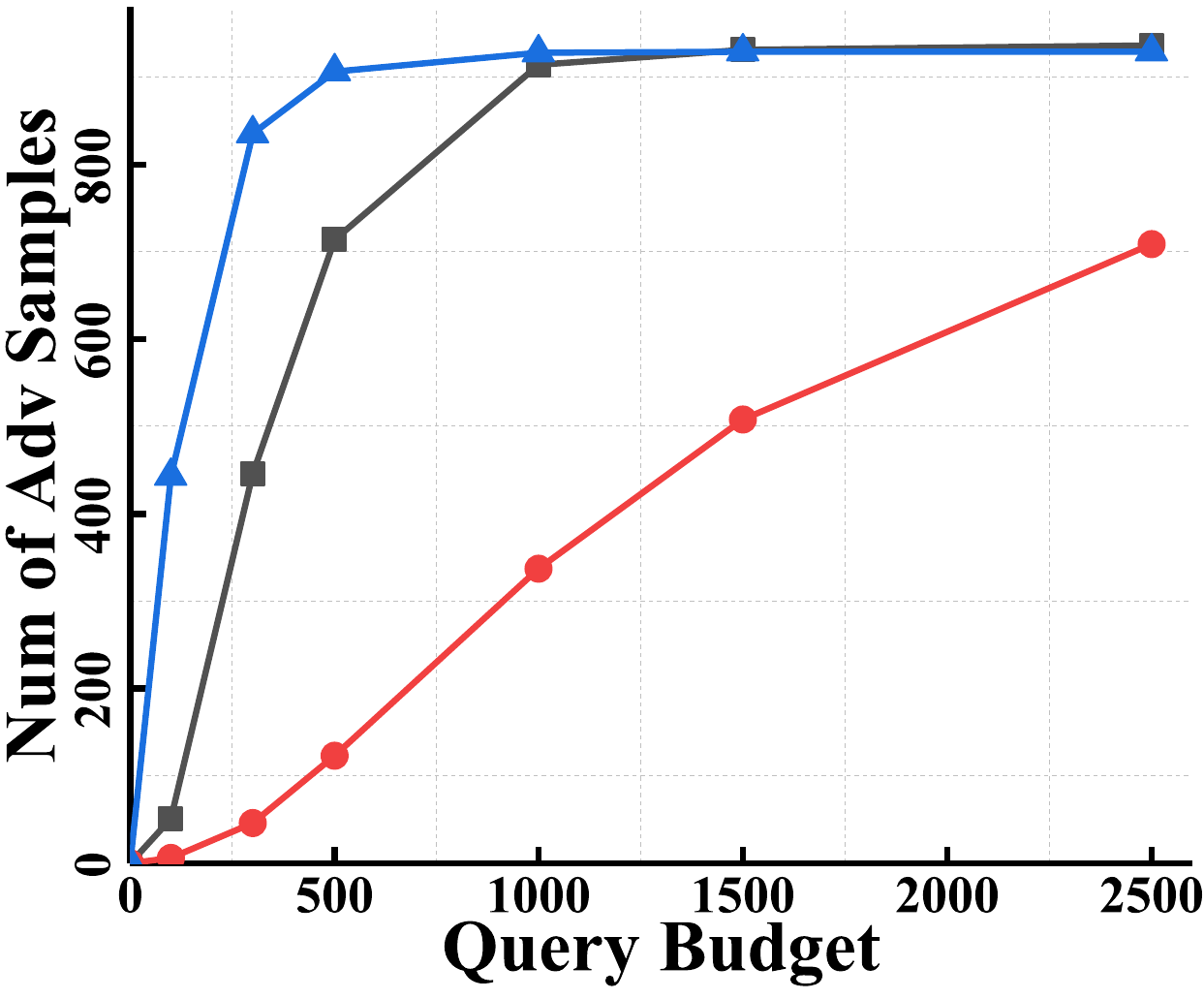}
    \end{minipage}
    \label{figure.yelp_budget_cnn}
}
\subfigure[\imdb{}]
{
    \begin{minipage}[b]{.3\linewidth}
        \centering
        \includegraphics[scale=0.245]{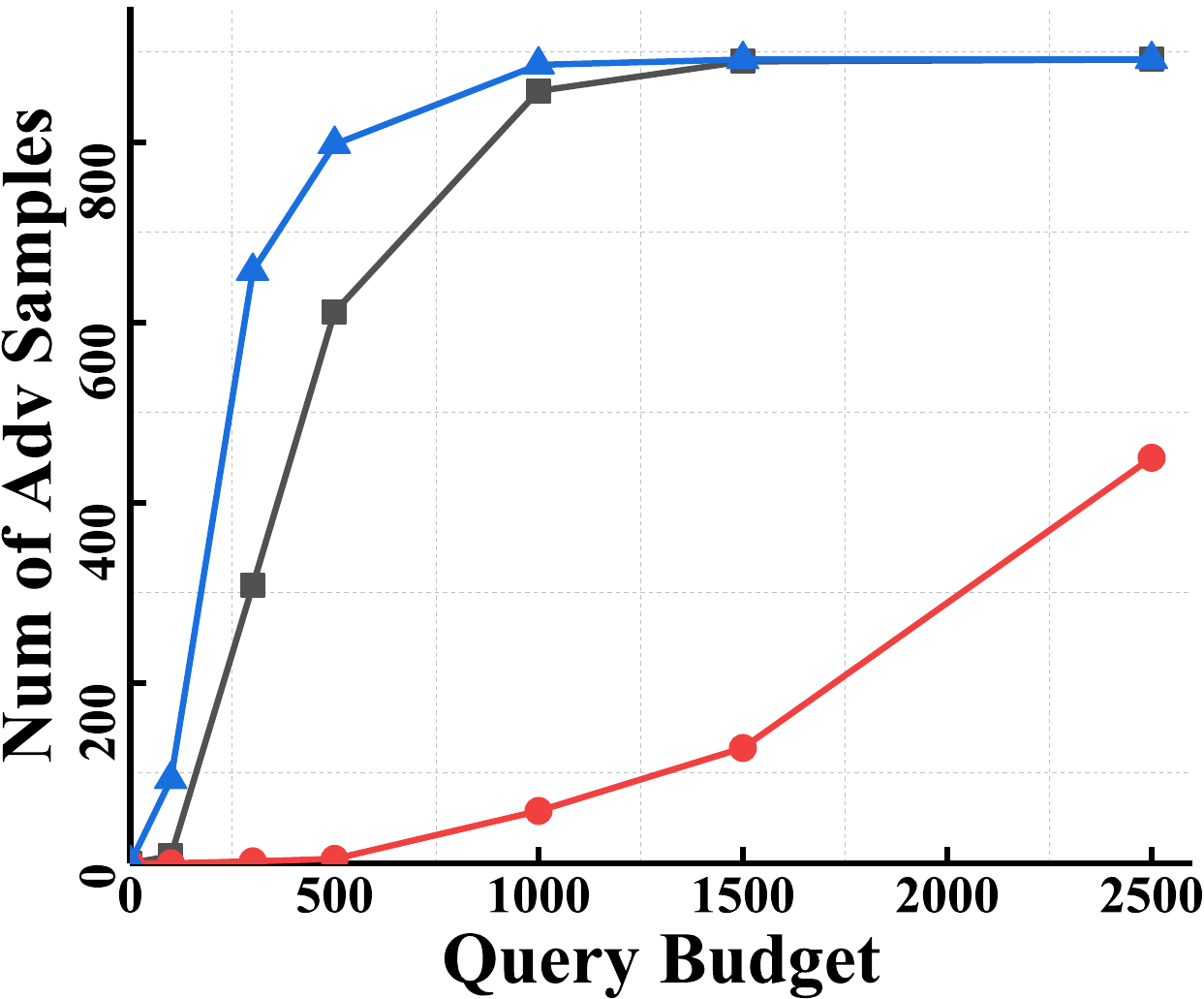}
    \end{minipage}
    \label{figure.imdb_budget_cnn}
}
\caption{The number of successful adversarial samples generated under varying the query budget $Q_\text{max}$ on WordCNN.
}\label{figure.budget_cnn}
\end{figure*}